\documentclass[journal]{IEEEtai}

\usepackage[colorlinks,urlcolor=black,linkcolor=blue,citecolor=blue]{hyperref}
\usepackage{color,array}
\usepackage{xcolor}
\usepackage{biblatex}
\usepackage{stfloats}
\usepackage{authblk}
\usepackage{graphicx}
\usepackage{amsmath}
\usepackage[justification=centering]{caption}
\usepackage[skins]{tcolorbox}
\usepackage[loadonly]{enumitem}

\newlist{arrowlist}{itemize}{1}
\setlist[arrowlist]{label=\(\rightarrow\),  
                    leftmargin=*,          
                    itemsep=0pt}           

\newtcolorbox{examplebox}{
  colback=gray!12,
  colframe=gray!50,
  boxrule=0.3pt,
  arc=2pt,
  left=6pt,right=6pt,top=4pt,bottom=4pt
}

\begin{document}

\title{\vspace{10mm} Mechanistic Interpretability in the Presence of Architectural Obfuscation \vspace{10mm}} 

\author{
\centering
\begin{minipage}{0.4\textwidth} \centering
    \textbf{Marcos Florencio*} \\
    \scriptsize INTELI – Institute of Technology and Leadership \\
    \vspace{0.5em}
    \texttt{marcos.florencio@sou.inteli.edu.br}
\end{minipage}
\hspace{2em} 
\begin{minipage}{0.4\textwidth} \centering
    \textbf{Thomas Barton*} \\
    \scriptsize INTELI – Institute of Technology and Leadership \\
    \vspace{0.5em}
    \texttt{thomas.barton@sou.inteli.edu.br}
\end{minipage}

\thanks{Marcos Florencio and Thomas Barton are undergraduate students in Software Engineering and Computer Science, respectively, graduating in 2025. This paper is part of an ongoing effort developed within the academic research path at INTELI – Institute of Technology and Leadership. Source code is available at \url{https://github.com/themarcosf/mech-interp-paper}.}}

\maketitle

\begin{abstract}
Architectural obfuscation—e.g., permuting hidden-state tensors, linearly transforming embedding tables, or remapping tokens—has recently gained traction as a lightweight substitute for heavyweight cryptography in privacy-preserving large-language-model (LLM) inference.  While recent work has shown that these techniques can be broken under dedicated reconstruction attacks, their impact on mechanistic interpretability has not been systematically studied.  In particular, it remains unclear whether scrambling a network’s internal representations truly thwarts efforts to understand how the model works, or simply relocates the same circuits to an unfamiliar coordinate system.

We address this gap by analyzing a GPT-2-small model trained from scratch with a representative obfuscation map. Assuming the obfuscation map is private and the original basis is hidden (mirroring an honest-but-curious server), we apply logit-lens attribution, causal path-patching, and attention-head ablation to locate and manipulate known circuits.

Our findings reveal that obfuscation dramatically alters activation patterns within attention heads yet preserves the layer-wise computational graph.  This disconnect hampers reverse-engineering of user prompts: causal traces lose their alignment with baseline semantics, and token-level logit attributions become too noisy to reconstruct.  At the same time, feed-forward and residual pathways remain functionally intact, suggesting that obfuscation degrades fine-grained interpretability without compromising top-level task performance.

These results establish quantitative evidence that architectural obfuscation can simultaneously (i) retain global model behaviour and (ii) impede mechanistic analyses of user-specific content.  By mapping where interpretability breaks down, our study provides guidance for future privacy defences and for robustness-aware interpretability tooling.
\end{abstract}

\vspace{\baselineskip}

\section{Introduction}
\vspace{\baselineskip}

\IEEEPARstart{A}{rtificial} Intelligence (AI) has become one of the most promising areas of computer science, exerting a profound impact in both academic and industrial settings thanks to rapid advances in Machine Learning (ML) and Deep Learning (DL) \cite{tjoa2020survey}. Defined as the ability of machines to simulate human cognitive processes \cite{lee2017deep}, AI is now embedded in countless real-world applications.  Artificial Neural Networks (ANNs), abstracting principles of biological cognition, underpin this progress by solving classification, prediction and pattern-recognition problems once thought intractable \cite{thakur2021fundamentals}.

Within this landscape, the Transformer architecture has emerged as a disruptive force, displacing recurrent and convolutional models through its self-attention mechanism \cite{rothman2024transformers, bengesi2024advancements, elhage2021mathematical}.  Flagship families, such as GPT, now dominate natural-language processing and are steadily migrating into vision, speech and multi-modal domains, processing ever-larger parameter counts that already number in the hundreds of billions \cite{campino2024unleashing}.  Their predictive power, however, comes at the cost of opacity: Transformers behave as “black boxes,” making it difficult to trace how inputs are transformed into outputs—an obstacle that is increasingly problematic in safety-critical domains such as healthcare, transportation and finance \cite{oviedo2022interpretable, wang2022interpretability, yu2021deep}.

To restore trust, two complementary research threads have flourished.  Mechanistic interpretability seeks to reverse-engineer the learned algorithms inside large models, decomposing attention heads and MLP pathways into human-readable circuits \cite{wen2022interpretability, wang2022interpretability}.  In parallel, secure inference aims to protect private prompts and proprietary weights when models are hosted by third parties, a practical necessity given the hardware demands of state-of-the-art Large Language Models (LLMs) \cite{thomas2025attack, oseni2021security}.  

Traditional cryptographic defences—secure multi-party computation and homomorphic encryption—offer strong guarantees but impose prohibitive latency and bandwidth overheads.  Consequently, lighter-weight obfuscation strategies have gained popularity: they attempt to scramble internal representations or token identities so that a curious server (or client) cannot easily recover the original text \cite{lin2024inversion, conti2022obfuscation}.

Large-scale LLM research is experimenting the feasibility of architectural obfuscation techniques, such as permutation-based schemes \cite{zheng2024permllm, yuan2024stip}, which shuffle hidden-state tensors across sequence or feature dimensions, embedding-matrix transforms such as glide reflections \cite{mishra2023sentinellm}, which randomise token vectors, and token-level manipulations and proxy prompts \cite{luo2024centaur}, which disguise the textual interface itself. These approaches promise privacy with minimal runtime cost and have become the dominant “lightweight” alternatives to heavy cryptography.

Simultaneously, a growing body of work highlights their fragility.  A linear-time matching attack recovers $\approx 100 \%$ of prompts from permuted hidden states \cite{lin2024ednn}, while the algorithm reconstructs entire vocabularies after glide-reflection obfuscation, effectively nullifying the intended secrecy \cite{luo2024centaur}. These findings suggest that many current defences merely rearrange information rather than hide it.

From a mechanistic point of view, the key question is not whether these techniques satisfy cryptographic definitions of privacy, but how much of the model’s internal computation remains intelligible once such obfuscations are applied.  Because permutations and linear transforms preserve algebraic structure, the causal circuits that copy subject names, route factual cues or compute logits might persist—merely expressed in a shuffled coordinate system.  Quantifying the extent to which these circuits survive offers a principled way to gauge the depth of architectural obfuscation beyond surface-level security claims.

This work addresses exactly that gap.  We do not propose new security proofs nor audit the privacy guarantees of existing schemes.  Instead, we ask: Can we still locate, characterise and manipulate the mechanisms that drive model behaviour once the observable representations have been permuted or otherwise obfuscated?

To answer this, we trained GPT-2-small under a representative obfuscation regime and applied state-of-the-art circuit-finding tools, thereby mapping which interpretability probes remain effective and where they fail.  Our results establish an empirical upper bound on the mechanistic opacity achievable with current permutation- and transformation-based defences, thereby informing both the interpretability community and designers of future privacy mechanisms.

\vspace{\baselineskip}
\section{Background}
\vspace{\baselineskip}

Dario Amodei’s recent essay "The Urgency of Interpretability" crystallizes a view that has been gaining ground across industry and academia: understanding what large models are “thinking” is no longer optional but a prerequisite for safe deployment \cite{amodei2024urgency}.  Amodei traces this imperative back to the systematic programme that Chris Olah began at Google Brain and later advanced at OpenAI.  Olah’s early vision work showed that deep networks contain individual units whose activations align with human concepts such as “wheel” or “car,” mirroring the neuroscientific idea of a “Jennifer Aniston neuron” \cite{quiroga2005aniston}.  In 2021, a Distill investigation extended this line of inquiry to multimodal neurons that respond jointly to text and image cues, illustrating that concept-level structure survives even in very large, cross-modal settings \cite{olah2021multimodal}.

When Olah and Amodei co-founded Anthropic, they shifted the spotlight from vision to language and laid down a formal blueprint for reasoning about Transformers.  \cite{elhage2021mathematical} introduced algebraic tools—logit-lens factorisations, residual-stream decompositions—and used them to expose canonical mechanisms such as induction heads, which copy earlier tokens, and softmax linear units, single neurons that act as token detectors \cite{elhage2021mathematical}.  These studies revealed that some neurons are directly interpretable but most are not; instead, they participate in dense superpositions where hundreds of features share the same vector space.  Superposition allows the model to encode more concepts than it has neurons, but it also scrambles human legibility \cite{olah2022superposition}.

Progress stalled until interpretability researchers imported sparse-coding ideas from signal processing. \cite{nanda2023towards} showed that dictionary-learning and sparse autoencoders can untangle superpositions into monosemantic features—linear subspaces whose activations correspond to crisp linguistic or stylistic concepts.  \cite{cunningham2023sparse} reached a similar conclusion independently, demonstrating that sparse autoencoders recover highly interpretable features even in production-scale GPT variants. \cite{anthropic2024mapping,anthropic2024scaling} then scaled the technique, mapping more than thirty million features in a mid-sized commercial model and introducing autointerpretability, a bootstrapped procedure in which the model itself generates natural-language descriptions of the features it houses.  Feature-level control is already practical: by amplifying a single “Golden Gate Bridge” vector, engineers produced “Golden Gate Claude,” a model that obsessively mentioned the bridge in otherwise unrelated contexts \cite{amodei2024urgency}.

The frontier has now moved from isolated features to circuits—structured constellations of features that implement multi-step reasoning.  In \cite{anthropic2025biology}, researchers traced how a prompt about “the state containing Dallas” activated a Dallas feature, routed through a located-in circuit to fire Texas, and finally triggered Austin under a capital-of circuit.  Although only a handful of such circuits have been located manually, the work hints that a full model might harbour millions of interacting sub-programs, each susceptible to targeted intervention once identified.

While mechanistic interpretability has marched ahead, privacy concerns have pushed practitioners toward architectural obfuscation as a lightweight defence for cloud-hosted inference.  Permutation-based schemes (\cite{zheng2024permllm, yuan2024stip}) randomly reorder hidden-state tensors, \cite{luo2024centaur} extends the idea with two-dimensional shuffles and masks, and glide-reflection transforms apply secret linear maps to the embedding matrix \cite{mishra2023sentinellm}.  These techniques claim to protect user prompts at wide-area-network latency, avoiding the severe overheads of homomorphic encryption or secure multi-party computation.  Yet dedicated reconstruction attacks paint a mixed picture: \cite{zheng2024permllm} reverse hidden-state permutations with near-perfect accuracy, and \cite{lin2024ednn} recover the entire vocabulary after glide-reflection using the EDNN algorithm.  Even if these results undermine privacy guarantees, they leave open a different and crucial question inspired by Amodei’s essay: do such transformations impede our ability to locate and manipulate the model’s internal mechanisms?

Answering that question is the purpose of this paper.  Mechanistic interpretability has supplied tools for dissecting unfettered models; the security community has catalogued ways to obfuscate them.  What is missing is a systematic examination of interpretability under obfuscation.  By analysing GPT-2-small protected with representative permutation and linear-transform schemes, and by measuring how logit-lens attribution, path patching, and circuit tracing survive or fail under each transformation, we aim to quantify the interpretability cost—or resilience—of today’s lightweight privacy defences.

\vspace{\baselineskip}
\section{Methodology}
\vspace{\baselineskip}

This section outlines the methodology adopted to implement and evaluate the proposed obfuscation technique applied to a transformer-based language model. The objective was to adapt an existing large language model architecture to incorporate client-specific encoding mechanisms while maintaining efficiency and performance during training and inference. The methodology encompasses the design of the model architecture, training data selection, optimization procedures, and evaluation strategies. The implementation of the seeded tokenizer is described in detail, the training process using the Fineweb-Edu dataset, the architectural modifications applied to the Transformer model, and the experimental setup employed to assess the model's capabilities.

\subsection{Seeded Tokenizer}
\vspace{\baselineskip}

The seeded tokenizer follows the implementation proposed by \cite{zheng2024permllm}, repackaged as a proof-of-concept tool suitable for experimentation in an academic setting.
The original implementation shows that if the elements entering every expensive non-linear block are first shuffled by a secret, fresh permutation, the model can compute non-linearities locally in plaintext; afterwards the parties run a second secure shuffle to map results back. Because the permutation lives only in secret shares, the model holder never recovers the true order, yet the cost of cryptographic-based solutions is avoided . The paper stresses that the security of this shortcut hinges on the permutation space, with larger spaces making brute-forcing infeasible in practice.

Our implementation distills the original to the level of the tokenizer. It first loads GPT-2’s vocabulary with Hugging Face, then draws a uniformly random permutation of the integers, thus creating a deterministic seeded mapping that re-indexes every token. This mapping is stored in a local cache so it can be used in all subsequent experiments. In effect, any text encoded through this tokenizer is now expressed in a permuted space, replicating the insight that a shuffled representation leaks far less direct information while remaining functionally equivalent once the model’s embedding and output layers are reordered accordingly.

\cite{zheng2024permllm} regenerates a fresh, hidden permutation for every forward pass and intertwines it with additive secret sharing, so that neither party ever learns the other’s weights or prompts; its correctness relies on running two secure permutation protocols per non-linear call and homomorphic dot-products at the output. The seeded-tokenizer code instead fixes a single permutation at startup and assumes whoever possesses the mapping is authorized to decode it.

Thus, this lightweight mechanism emulates the permutation shield idea, allowing the measurement of its impact on model-quality degradation when embeddings are re-indexed.

\subsection{Model Architecture}
\vspace{\baselineskip}

In order to test the obfuscation technique, a custom version of the GPT-2-small model, with 124 million parameters, was trained \cite{radford2019language}. The GPT-2 model is a large-scale language model developed by OpenAI, which has been widely used for various natural language processing tasks. The custom version of the model was modified to include a pre-seeded tokenizer that transforms input data into client-specific formats, allowing for further encryption and obfuscation of the data during inference.

The dataset used to train the original GPT-2 is not publicly available, but \cite{radford2019language} mentions that in order to build the training dataset, ``all outbound links from Reddit, a social media platform, which received at least 3 karma'' were scraped, totalling 45 million links and 40 GB of text. By contrast, GPT-3 was trained using a mixture of data from Common Crawl, WebText2, books, and Wikipedia \cite{brown2020language}. For our case, the custom model was trained on a large corpus of text data called Fineweb-Edu, available on HuggingFace, which consists of 10.0 billion tokens extracted from educational content \cite{penedo2024fineweb}. The training process involved optimizing the model's parameters to minimize the difference between the predicted and actual outputs, allowing it to generate coherent and contextually relevant text.

It is worth noting that GPT-2 implements a modified version of the Transformer model introduced in \cite{vaswani2017attention}. In particular, it does not have the encoder part of the Transformer model, only the decoder part. The decoder consists of a stack of $N=12$ identical layers. Each layer has two sub-layers: a multi-head self-attention mechanism and a position-wise fully connected feed-forward network. Other major departures from the original Transformer, as described in Section 2.3 of \cite{radford2019language}, include moving layer normalization to the input of each sub-block—similar to a pre-activation residual network—and adding an additional layer normalization after the final self-attention block. An important consequence of moving the layer normalization to the input of each sub-block is that the normalization is then applied outside of the residual stream, improving the stability of the residual path during training.

Additionally, we employ Flash Attention \cite{dao2022flashattention}, a memory-efficient attention mechanism that reduces the memory footprint of the model during training and inference through a kernel-fusion operation, condensing four different operations into a single fused kernel. To achieve this, it relies on an algorithmic rewrite of the softmax function, previously proposed by \cite{milakov2018online}, which computes the softmax function with fewer memory accesses, allowing for faster and more efficient training of large-scale language models, enabling efficient training on limited hardware resources.

\subsection{Activation Function}
\vspace{\baselineskip}

The activation function used in the feed-forward network is the GELU (Gaussian Error Linear Units) function. The GELU function is a smoothed approximation of the rectifier linear unit (ReLU) function, as proposed by \cite{hendrycks2016gaussian}. The GELU function is defined as:

\[
\text{GELU}(x) = x \cdot \Phi(x) = x \cdot \frac{1}{2} \left(1 + \text{erf}\left(\frac{x}{\sqrt{2}}\right)\right)
\]

where $\Phi(x)$ is the cumulative distribution function of the standard normal distribution, and $\text{erf}(x)$ is the error function. The GELU function has been shown to outperform other activation functions, such as ReLU and $\tanh$, in various deep learning tasks.

\subsection{Parameter Sharing: Token Embeddings and Output Layer}
\label{subsec:param_sharing}
\vspace{\baselineskip}

In the Transformer model, the authors propose sharing parameters between the input and output layers by reusing the same weight matrix between the embedding layers and the pre-softmax linear transformation, following the approach of \cite{press2016using}. They propose that using the same embedding matrix for both the input and output layers reduces the number of parameters and improves generalization: ``We call $U$ the input embedding, and $V$ the output embedding. In both matrices, we expect rows that correspond to similar words to be similar: for the input embedding, we would like the network to react similarly to synonyms, while in the output embedding, we would like the scores of words that are interchangeable to be similar.''

\subsection{Weight Initialization}
\vspace{\baselineskip}

In the code released by OpenAI \cite{radford2019language}, the weights are initialized with a standard deviation of 0.02, and the biases are initialized to zero. The token embeddings are also initialized with a standard deviation of 0.02, while the positional embeddings are initialized to 0.01. This is consistent with the recommendations from the original Transformer paper, which suggests initializing the weights with a normal distribution with a standard deviation of $\sqrt{2 / (n_{\text{in}} + n_{\text{out}})}$, where $n_{\text{in}}$ and $n_{\text{out}}$ are the number of input and output units, respectively \cite{vaswani2017attention}.

Additionally, \cite{radford2019language} mentions using a modified initialization that accounts for residual path accumulation with model depth. Specifically, the weights of residual layers are scaled at initialization by a factor of $1/\sqrt{N}$, where $N$ is the number of residual layers.

\subsection{Hardware and Training Schedule}
\vspace{\baselineskip}

The custom model was trained on four parallel NVIDIA Tesla L4 GPUs, each with 24 GB of memory, using the DDP (Distributed Data Parallel) module available in the PyTorch framework \cite{pytorchddp}. DDP provides data parallelism by synchronizing gradients across each model replica. The optimization algorithm used was AdamW, a variant of Adam that incorporates weight decay and momentum buffers, which helps stabilize the training and improve convergence \cite{brown2020language}. The optimizer was configured with $\beta_1 = 0.9$, $\beta_2 = 0.95$, and $\epsilon = 10^{-8}$, with gradient clipping set to a maximum norm of 1.0 to prevent exploding gradients.

Training used a batch size of 8 samples without replacement until epoch boundaries to minimize overfitting. The initial learning rate was $6 \times 10^{-4}$, subject to a cosine decay schedule with 10\% warmup, following the guidelines from \cite{brown2020language}. Validation loss was used to monitor progress, and checkpoints were saved at regular intervals.

To supplement the validation set, HellaSwag was used to evaluate model performance on a more challenging commonsense reasoning task \cite{zellers2019hellaswag}. HellaSwag is a benchmark of 10,042 multiple-choice questions with four answer options. The model was evaluated using token probabilities to compute the cross-entropy loss for each answer. The original GPT-2-small model achieved a score of 0.2955, which was used as a baseline for determining the stopping criteria during training.

\vspace{\baselineskip}
\section{Results and discussion}
\vspace{\baselineskip}

This section presents the empirical core of our study, contrasting how a baseline GPT-2 small model and its obfuscated counterpart handle a controlled anaphora-tracking task, first proposed by \cite{wang2022interpretability}. We first formalize the task, then quantify each model’s ability to single out the correct indirect object by inspecting the logit gap between competing name tokens. These results anchor a deeper mechanistic analysis—spanning residual stream, component and head patching—that illuminates where and how obfuscation alters the network’s internal representations while preserving output-level performance.

\subsection{Task Definition}
\vspace{\baselineskip}

The task is to identify the indirect object in a sentence, given a subject and an indirect object, as proposed in \cite{wang2022interpretability}, but using two different versions of the same model: the standard GPT-2 small model and a custom version trained on obfuscated data. Each prompt is given twice - one with the first name as the indirect object, one with the second name. These prompts are composed only of single-token names and the corresponding names are always in the same token positions. For example:

\begin{examplebox}
  \begin{arrowlist}
    \item When John and Mary went to the shops, John gave the bag to Mary.
    \item When Tom and James went to the park, Tom gave the ball to James.
    \item When Dan and Sid went to the shops, Dan gave an apple to Sid.
    \item After Martin and Amy went to the park, Martin gave a book to Amy.
  \end{arrowlist}
\end{examplebox}

This task is meaningful because attention is especially apt at primitive operations such as looking nearby or copying information. For example, a simple model could learn a skip trigram like "John...to \textrightarrow{} John". However, distinguishing between multiple mentions of the same name requires learning a more complex internal representation.

The model must learn to predict \texttt{Mary} and not \texttt{John}. For that, it needs a head that attends to all previous names but inhibits duplicates. This head then contributes to the logits in favor of the correct indirect object.

The symmetry can be broken at the second mention of \texttt{John}. Hence, one might expect a head to detect duplicate tokens and another to move that information to the token after \texttt{to}.

Because token positions vary across inputs, language models use early and late layers to convert from token-level representations to internal representations and back, as discussed in Section \ref{subsec:param_sharing}. This will be relevant when analyzing early attention heads acting as duplicate token detectors.

The evaluation metric is the logit difference between the indirect object and the subject. Since the softmax is translation invariant, we can center the logits. Therefore, interpreting output logits reduces to projecting the residual stream onto the difference vector between the two unembedding directions.

Furthermore, for interpretability, the product $W_Q W_K^T$ is analyzed using SVD: $W_Q W_K^T = USV^T$. Then $W_Q = U\sqrt{S}$ and $W_K = V\sqrt{S}$, yielding symmetric and orthogonal representations for keys and queries. The same applies to $W_V$ and $W_O$.

\subsection{Logit Attribution}
\label{subsec:logit_attr}
\vspace{\baselineskip}

At the heart of transformer-based architectures lies the residual stream, a central computational structure that accumulates and propagates information throughout the network. Rather than passing activations solely from one layer to the next, transformers use residual connections to aggregate outputs from multiple layers, effectively building a running summary of all transformations applied to the input sequence. This stream is modified at each layer by the outputs of both attention heads and MLP blocks, and it ultimately determines the model’s predictions \cite{elhage2021mathematical}.

In the final stage of the forward pass, the model produces token-level logits by projecting the final residual stream through the unembedding matrix, denoted as $W_U$, which maps internal representations back into the vocabulary space. The model’s output probability distribution is then computed via the softmax function applied to these logits.

To assess how the model differentiates between potential output tokens, the logit difference between the correct and incorrect choices is analyzed. This difference serves as a proxy for the model’s confidence and decision-making process. Formally, if $\vec{x}$ represents the residual stream at the final token position,
$\vec{L}$ represents the log probabilities of the outputs, and $\vec{p}$ represents the probabilities of the outputs after the aplication of the softmax, then the following relations hold:

$$
p_i = \mathrm{softmax}(\vec{x})_i = \frac{e^{x_i}}{\sum_{i=1}^{n} e^{x_i}}
$$

and

$$
L_i = \log(p_i)
$$

Combining these:

$$
L_i = \log \frac{e^{x_i}}{\sum_{j=1}^{n} e^{x_j}} = x_i - \log \sum_{j=1}^{n} e^{x_j}
$$

The sum term on the right is the same for all $i$:

$$
L_i - L_j = x_i - x_j
$$

In other words, the logit difference $x_i - x_j$ is the same as the log probability difference $L_i - L_j$, motivating the use of logit differences to understand the model's outputs, since getting an output logit is equivalent to projecting onto a direction in the residual stream.

If the final value in the residual stream for a single sequence and a position within that sequence is $\vec{x}$ (i.e., a vector of length $d_{\text{model}}$), then the logits are obtained by multiplying by the unembedding matrix $W_U$ (which has shape($d_{\text{model}}$, $d_{\text{vocab}}$)):

$$
\text{output} = \vec{x}^T W_U
$$

The \emph{logit difference direction (or vector)} between two tokens $i$ and $j$ is given by:

$$
\text{logit diff}_{ij}  = \vec{x}^T (W_U[:, i] - W_U[:, j])
$$

This means that the logit difference direction is the projection of the residual stream onto the vector $W_U[:, i] - W_U[:, j]$, because it points in the direction of the largest logit difference between the two tokens, capturing the axis along which the model must distinguish between the correct and incorrect answers. The scalar projection of the residual stream onto this vector quantifies the model’s preference for one token over the other.

Crucially, because the softmax operation is translation invariant (i.e., shifting all logits by the same scalar does not change the output probabilities), the only meaningful quantities are differences between logits. This allows us to center our interpretability analysis on directional contributions within the residual stream, rather than on absolute values.

To break down the origin of the final prediction, we employ logit attribution, a technique that decomposes the residual stream into additive contributions from every prior component—individual attention heads, MLPs, and residual connections at each layer. This decomposition enables us to identify which parts of the network architecture contribute most significantly to the correct prediction and how the computation is distributed temporally and hierarchically.

\begin{figure}[h]
    \centering
    \includegraphics[width=0.45\textwidth]{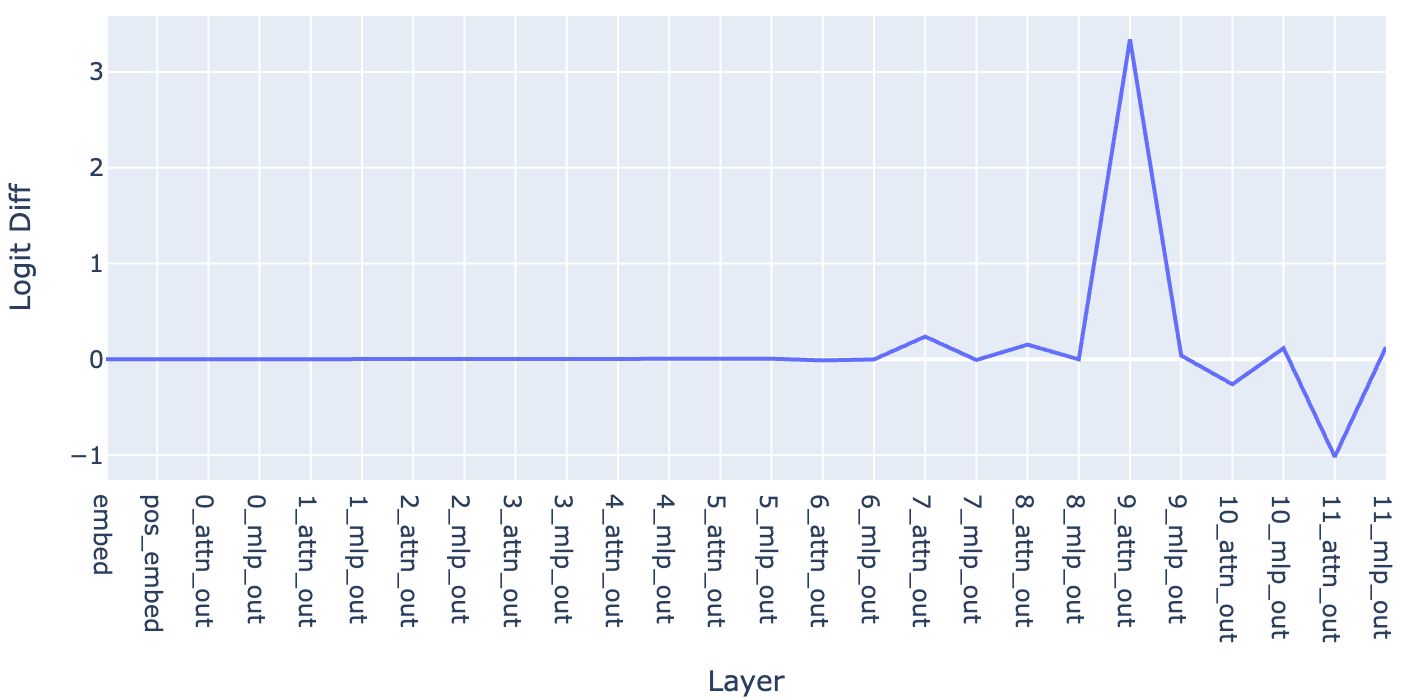}
    \caption{Logit difference - by layer}
    \label{fig:layer_logit_diff}
\end{figure}

The empirical results (as visualized in Figure \ref{fig:layer_logit_diff}) reveal that the model’s ability to perform the task is not uniformly distributed across layers. Notably:

\begin{itemize}
    \item Layers 1 through 6 exhibit minimal or noisy contributions to the final decision.
    \item Layer 7 marks the onset of meaningful computation.
    \item Layer 9 stands out as the most influential, generating the highest positive logit difference.
    \item Surprisingly, layers 10 and 11 contribute negatively, actively decreasing the logit gap and potentially introducing confusion or conflicting signals.
\end{itemize}

This non-monotonic behavior underscores a key insight in transformer interpretability: deeper layers do not necessarily correspond to better or more refined understanding. Some layers may encode auxiliary heuristics or capture patterns that, while useful for general language modeling, interfere with specific task-oriented reasoning.

\begin{figure}[h]
    \centering
    \includegraphics[width=0.45\textwidth]{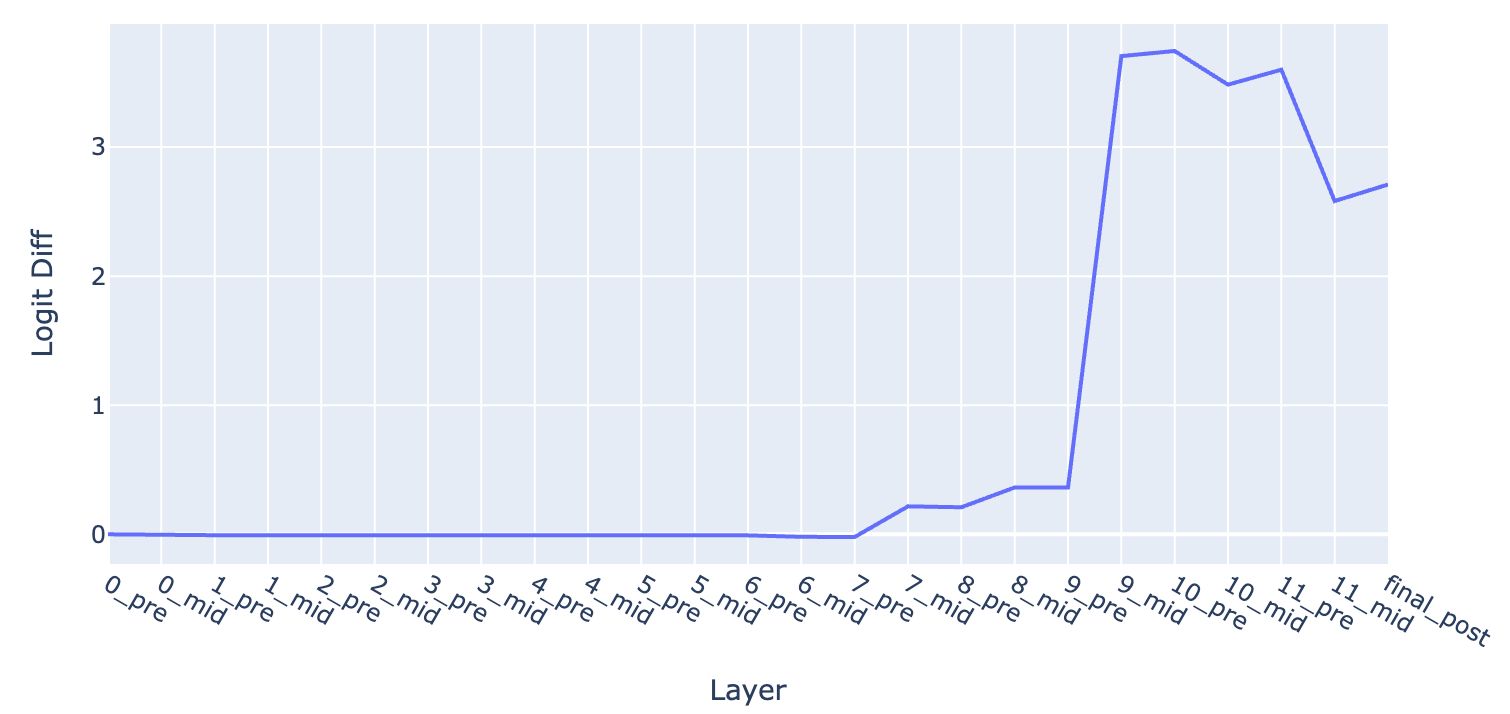}
    \caption{Logit difference - accumulated}
    \label{fig:head_logit_diff}
\end{figure}

Drilling down into the residual stream further, we examine attention head-level contributions, as depicted in Figure~\ref{fig:head_logit_diff}. This granularity reveals that the model’s predictive power is not evenly spread across all attention heads but is instead concentrated in a few specialized units. Specifically:

\begin{itemize}
    \item Heads 9.6 and 9.9 are identified as the primary positive contributors, suggesting they are central to the task-specific computation—likely resolving referential ambiguity or transferring relevant information between subject and indirect object tokens.
    \item In contrast, heads 10.7 and 11.10 are detrimental, consistently reducing logit confidence in the correct answer. These may encode features orthogonal or even antagonistic to the task, possibly representing overfitting or misalignment between general and specific objectives.
\end{itemize}

This division of labor among attention heads emphasizes the transformer’s modularity: individual components learn to specialize in different types of processing, and only a subset becomes critical for any given task.

A particularly illuminating finding emerges when comparing these results with those from the obfuscated model, which was trained to operate on transformed input representations for privacy-preserving inference. Remarkably, the overall pattern of head behavior remains broadly consistent. The same heads that dominate decision-making in the standard model continue to exert similar influence in the obfuscated variant. This suggests that the high-level circuit structure—i.e., the flow of information through the model—remains largely intact despite significant transformations to the input representation.

However, at the neuron-level, internal activations do diverge. That is, while the macro-level functional roles of heads are preserved, the micro-level computations—the precise values, firing patterns, and internal representations—are altered. This phenomenon indicates that the task-specific circuitry is robust to input distortions and that transformers can adapt their internal representations to compensate for obfuscation, preserving performance and interpretability at a structural level even when fine-grained details shift.

Note that in both Figures \ref{fig:layer_logit_diff} and \ref{fig:head_logit_diff} a layer is the \textit{kth} layer in a stack of transformer blocks, but each block consists of an attention layer and an MLP layer. The output of each attention layer is the result of the sum of the outputs of each attention head. In the GPT-2 model, each attention layer consists of 12 heads, which each act independently and additively. The standard way to compute the output of an attention layer is by concatenating the mixed values of each head, and multiplying by the output weight matrix. But, as described in \cite{elhage2021mathematical}, this is equivalent to splitting the output weight matrix into a per-head output and adding them up, including an overall bias term for the entire layer.

Image \ref{fig:image_3} below shows that only a few heads carry the weight of inference for the task - heads 9.6 and 9.9 contributing positively, explaining why attention layer 9 is so important, while heads 10.7 and 11.10 contribute a lot negatively, explaining why attention layer 10 and layer 11 are actively harmful. These correspond to some of the name movers and negative name movers discussed in [2]. There are also several heads that matter positively or negatively but less strongly.

\begin{figure}[h]
    \centering
    \includegraphics[width=0.45\textwidth]{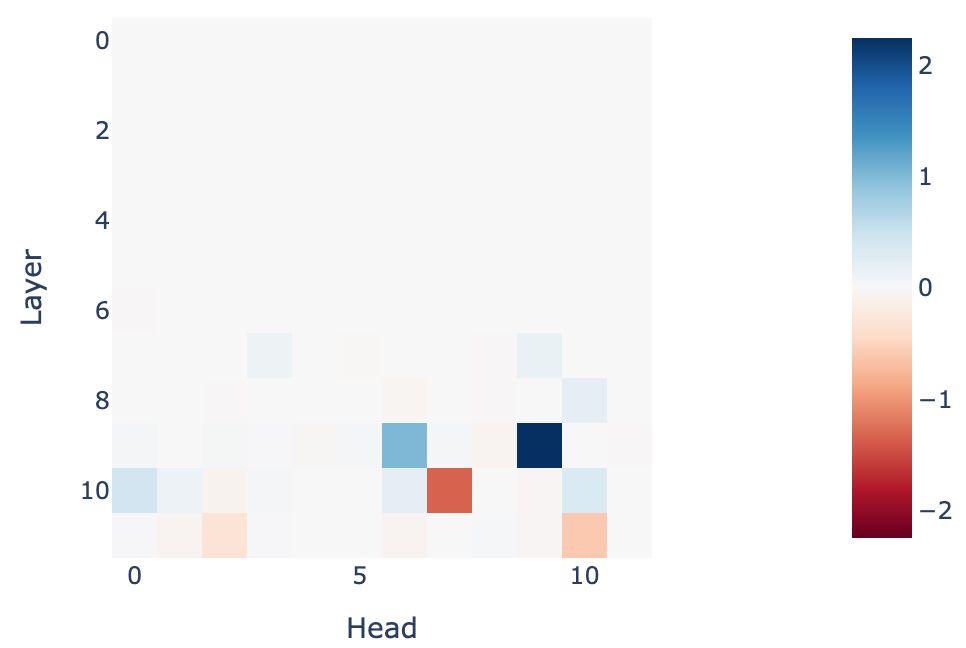}
    \caption{Logit difference - by head}
    \label{fig:image_3}
\end{figure}

In other architectures, it would be expected that a given computing unit would convey information about the token looked at, maybe accounting for the context of the token. But attention heads move information from the residual stream position corresponding to the input token. Especially later on in the model, there may even be components in the residual stream that have nothing to do with the input token, e.g., the period at the end of a sentence may contain summary information for that sentence, and the head may solely move that.

An interesting observation is that no difference can be seen between the two models in terms of overall behavior of attention heads and layers, which suggests that the obfuscation did not affect the model's ability to solve the task, and that the model is able to solve it in a way that is robust to obfuscation. Only by observing a level below the attention heads, at the neuron level, can differences be seen. This is because the obfuscation changed the way that the model's neurons are activated, but not the way that the model's attention heads are used to solve the task.

\subsection{Activation Patching}
\vspace{\baselineskip}

Activation patching is a diagnostic technique introduced in the mechanistic interpretability literature \cite{wang2022interpretability} for localizing the internal components of a neural network that are causally responsible for a given behavior. The goal is to identify where in the network and at what stage the model encodes the critical information necessary for solving a specific task.

The fundamental idea behind activation patching is to run the model under two contrasting conditions:

\begin{enumerate}
    \item A clean run, where the model receives an input known to yield the correct prediction, and
    \item A corrupted run, where the input is slightly altered (often syntactically or semantically) so that the model makes an incorrect prediction.
\end{enumerate}

After obtaining both forward passes, one iteratively replaces selected internal activations from the corrupted run with those from the clean run—at a specific layer and token position—then observes whether the model’s output improves. If substituting the clean activation into the corrupted context restores or improves performance (i.e., increases the logit difference in favor of the correct output), it suggests that the patched activation encodes information critical for the task.

Formally, we define the patching effect using the change in the logit difference (as discussed in Section \ref{subsec:logit_attr}) between the correct and incorrect answer tokens. The metric used to quantify the effect of a patch is the relative increase in logit difference caused by substituting in clean activations. This value provides a direct, interpretable measure of how much influence a particular component of the network has on the decision-making process.

\vspace{\baselineskip}
\textbf{Patching Paradigms}\\

Two complementary variants of activation patching exist, each answering a different causal question about internal representations:

\begin{itemize}
    \item \textbf{Denoising}: This involves running the model on a corrupted input (i.e., one that leads to an incorrect output), and selectively patching in activations from the clean run. This approach asks: “Is this clean activation sufficient to fix the model’s behavior?” If performance improves, then the patched activation is carrying information necessary for producing the correct output.
    \item \textbf{Noising}: This is the inverse experiment: the model is run on a clean input, and corrupted activations are patched in. This approach asks: “Is this component necessary for correct behavior?” If patching in corrupted activations degrades performance, it implies that the original activation was crucial to the success of the clean run.
\end{itemize}

While both approaches provide useful insights, denoising is generally preferred in mechanistic interpretability research, particularly in the early stages of circuit discovery. This is because it directly tests sufficiency: whether the presence of a given signal at a specific location is enough to restore correct behavior. By contrast, noising tests necessity, which can be harder to interpret—especially in cases where multiple redundant pathways exist in the model.

In practice, activation patching allows researchers to build causal maps of information flow within the network. By systematically testing all layer/position pairs, one can visualize which regions of the model are encoding, transforming, or propagating information relevant to a specific task or token prediction. This localization enables the discovery of functional circuits—clusters of attention heads and neurons whose coordinated activity gives rise to interpretable behavior.

In the context of our study, activation patching serves as a vital tool for comparing the base model with its obfuscated counterpart, shedding light on whether and how internal representations shift when the input is transformed for privacy. It allows us to determine not only whether the model still solves the task, but where in the architecture that solution is implemented, and how those solution pathways may reorganize under data obfuscation.

\vspace{\baselineskip}
\subsubsection{Residual Stream Patching}

To trace how information flows through the model during inference, we perform residual stream patching. This method targets the residual stream—the central vector representation that accumulates updates from each attention and MLP block—and systematically replaces the corrupted version of this stream with its clean counterpart. This is done at the start of each layer and at every token position, allowing us to identify where in the model’s computation graph task-relevant information is injected, transferred, or preserved.

\begin{figure}[h]
    \centering
    \includegraphics[width=0.4\textwidth]{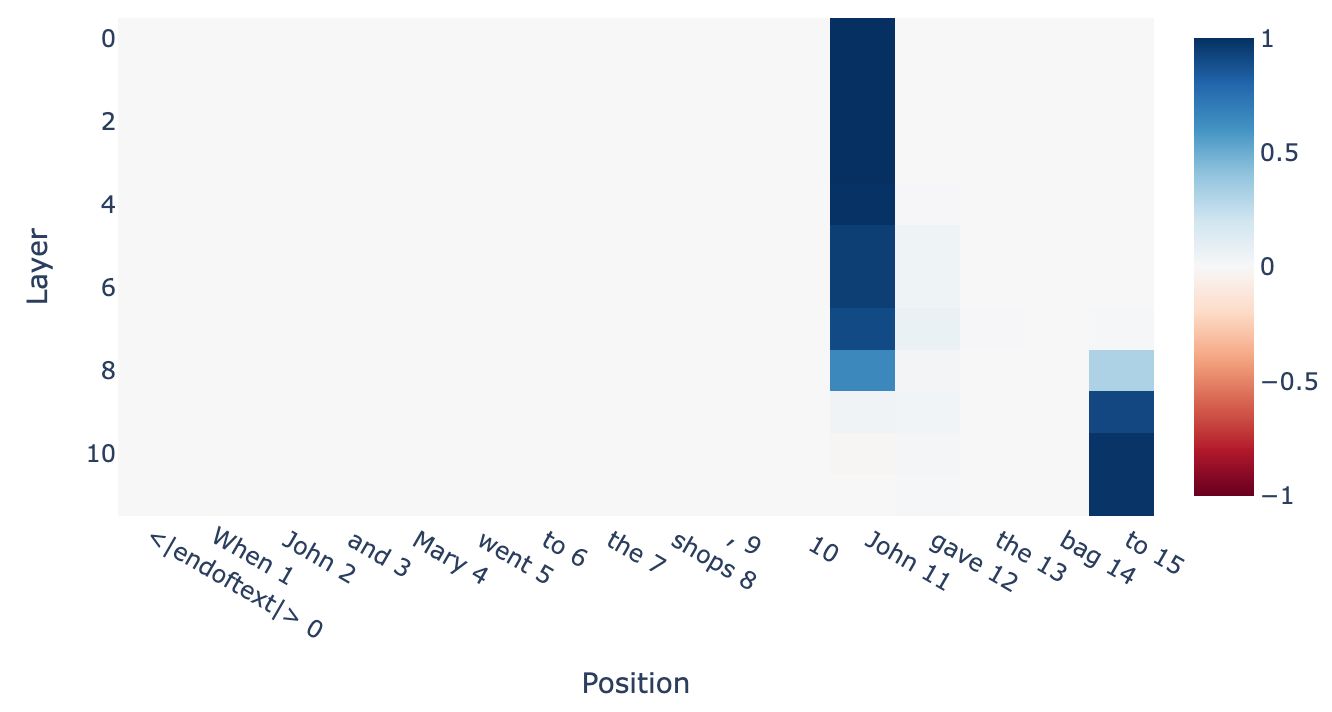}
    \caption{Residual stream patching - base model}
    \label{fig:image_7}
\end{figure}

For clarity, the visualization in Figure~\ref{fig:image_7} plots token positions on the x-axis (taken from a reference prompt such as: “When John and Mary went to the shops, John gave the bag to Mary”) and layer depth on the y-axis. Each patching experiment replaces the residual stream vector at a specific (layer, token) coordinate in the corrupted run with the corresponding vector from the clean run. The metric used is the relative change in logit difference, averaged over all eight prompts. However, token labels (e.g., “John”, “Mary”, “shops”) are derived from the first prompt for interpretability.

This analysis yields several key insights into the spatial and temporal structure of the model’s computation:

\begin{itemize}
    \item Computation is highly localized. Rather than being distributed across many tokens and layers, the model’s decision-making is concentrated at specific layer-token coordinates. This reveals the presence of a compact and focused subcircuit dedicated to solving the indirect object identification task.
	•	The most critical site of information encoding is the second subject token, corresponding to the indirect object (e.g., “Mary”). Early in the forward pass, this token stores distinguishing features necessary to resolve the recipient of the action.
    \item As the model progresses through the layers, this information is transferred to the final token, often the END token, where the decision is finalized. This reflects the model’s need to gather context from earlier mentions and consolidate it at the generation step, consistent with transformer-style autoregressive inference.
    \item Notably, the transition of relevant information occurs most prominently between layers 6 and 8. After layer 8, the model appears to have already completed the bulk of its reasoning. Layers 9 through 11, while still active, do not materially change the residual stream’s direction along the logit difference axis. Instead, they preserve and transport the already-computed representation with minimal interference. This behavior resembles identity transformations, where deeper layers merely relay useful information rather than transform it further.
\end{itemize}

This pattern highlights a useful interpretability property: transformers often complete meaningful sub-tasks early, with later layers acting more as memory buffers than as sites of further reasoning. In the context of indirect object resolution, the decisive computation is both early and traceable, reinforcing the feasibility of mechanistic reverse-engineering in this domain.

\begin{figure}[h]
    \centering
    \includegraphics[width=0.4\textwidth]{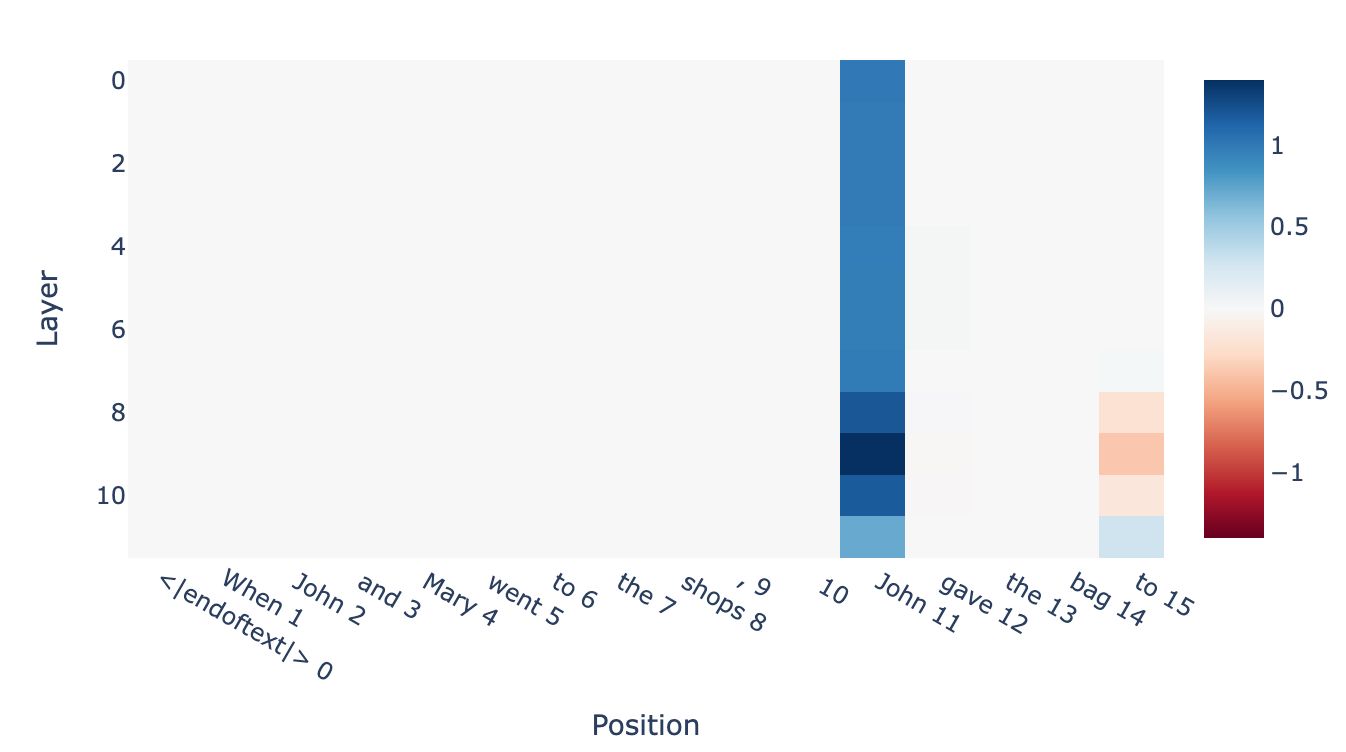}
    \caption{Residual stream patching - obfuscated model}
    \label{fig:image_8}
\end{figure}

When we apply the same patching procedure to the obfuscated version of the model (shown in Figure~\ref{fig:image_8}), a similar overarching structure emerges, but with notable differences in sharpness and stability:

\begin{itemize}
    \item The localization of computation remains. The second subject token still serves as the main site of initial information storage, and the transfer to the END token is still visible—suggesting that the overall algorithm implemented by the model is preserved even under obfuscated conditions.
    \item However, the signal becomes less sharp and more diffuse across both token positions and layers. This means the impact of patching is spread more broadly, implying that the model may be distributing computation over a larger number of heads or positions to compensate for the loss of clarity introduced by input transformation.
    \item Most critically, later layers in the obfuscated model remain active in modifying the residual stream. Unlike the base model, which stabilizes by layer 8, the obfuscated model continues to transform the internal representation in layers 9–11. This suggests that the obfuscated model requires additional computation in deeper layers to reconstitute or adapt the encrypted information into a usable format.
\end{itemize}

These observations support a key hypothesis: although high-level circuit structure is conserved under obfuscation, the burden of computation is redistributed. The model needs to work harder in later stages, indicating that obfuscation imposes a representational tax that the model pays via deeper processing.

In summary, residual stream patching reveals that:
\begin{itemize}
    \item The indirect object identification task is mediated by a small number of specific layers and token positions.
    \item In the base model, most relevant computation concludes by layer 8, after which information is merely preserved.
    \item In the obfuscated model, while the computational path remains recognizable, it becomes less efficient and more dispersed, especially in the final layers, which actively modify representations rather than passively retaining them.
\end{itemize}

This reinforces the notion that interpretability methods such as patching remain robust even under transformations aimed at enhancing privacy—and that the semantic flow of information through transformer models can survive significant architectural or input-level perturbations.

\subsubsection{Patching by Block Component}

While residual stream patching gives a high-level view of where important information is stored or transferred within the model, a more fine-grained analysis is necessary to disentangle the contributions of individual architectural components. To this end, we refine the patching procedure by intervening at three specific points within each transformer block: the input to the attention sublayer, the output of the attention sublayer, and the output of the MLP sublayer. This methodology allows us to probe not only where information is processed, but how it is distributed between attention mechanisms and feedforward transformations, offering a richer understanding of the model’s internal logic.

\begin{figure*}[t]
    \centering
    \includegraphics[width=\textwidth]{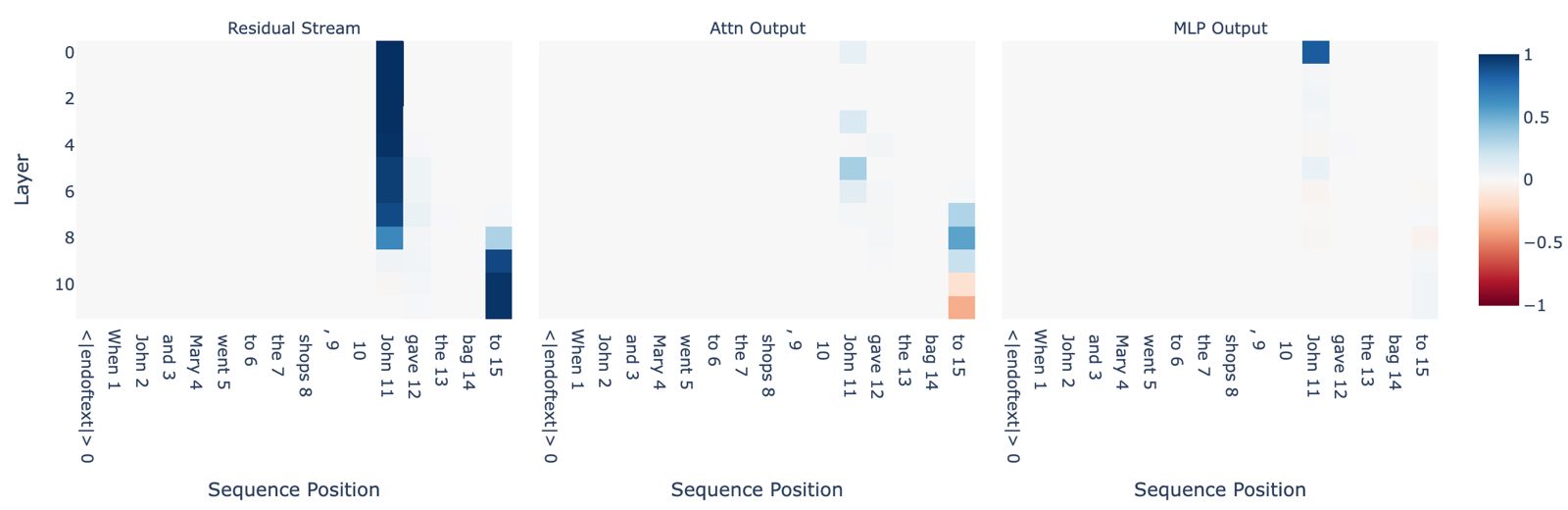}
    \caption{Component patching - base model}
    \label{fig:image_9}
\end{figure*}

The results for the base model (Figure~\ref{fig:image_9}) reveal a pattern of computation that is both sharply localized and semantically coherent. When patching activations before and after attention layers, it becomes evident that several attention layers play a critical role in solving the indirect object identification task. Specifically, early layers (up through layer 7) exhibit meaningful activity around the position of the second subject token, while later layers—especially layers 8 and 9—show significant contributions at the final token (typically END), where the prediction is generated. The model exhibits a striking focus: activations corresponding to all other token positions and layers are essentially inert, suggesting that only a narrow subset of the architecture is functionally engaged in performing the task.

This finding aligns closely with prior residual stream patching observations and further confirms that the model performs a deliberate and efficient computation. Initially, the relevant entity (the indirect object) is encoded at the point of its mention; later, the information is retrieved and moved to the generation site via specific attention heads. In terms of contribution polarity, the outputs of attention layers in layers 7, 8, and 9 are predominantly positive, reinforcing the notion that these heads are actively responsible for transporting meaningful semantic content across the sequence. In contrast, the outputs of layers 10 and 11 tend to contribute negatively or not at all, suggesting that these deeper layers either interfere with or do not meaningfully enhance the core reasoning path. This pattern supports the interpretation that layers 7 through 9 are the computational backbone of the indirect object circuit, and that deeper layers may serve secondary or even counterproductive roles in this context.

The behavior of the MLP sublayers stands in stark contrast. Across the board, MLPs show minimal impact on the model’s decision-making, with the single and notable exception of MLP0. This anomaly is consistent with prior observations across transformer models like GPT-2, where ablating MLP0 has been shown to severely degrade performance even on unrelated tasks. The prevailing interpretation is that MLP0 functions not as a task-specific processing unit, but rather as a kind of embedding augmenter—a critical stage in which raw token embeddings are enriched and made accessible to later layers. Consequently, even though the first attention layer performs relatively little logical processing, it benefits from an enhanced representational substrate courtesy of MLP0. Thus, the attention heads in later layers can effectively operate on a more structured and informative input, with the residual stream serving as the medium for this enriched signal.

\begin{figure*}[t]
    \centering
    \includegraphics[width=\textwidth]{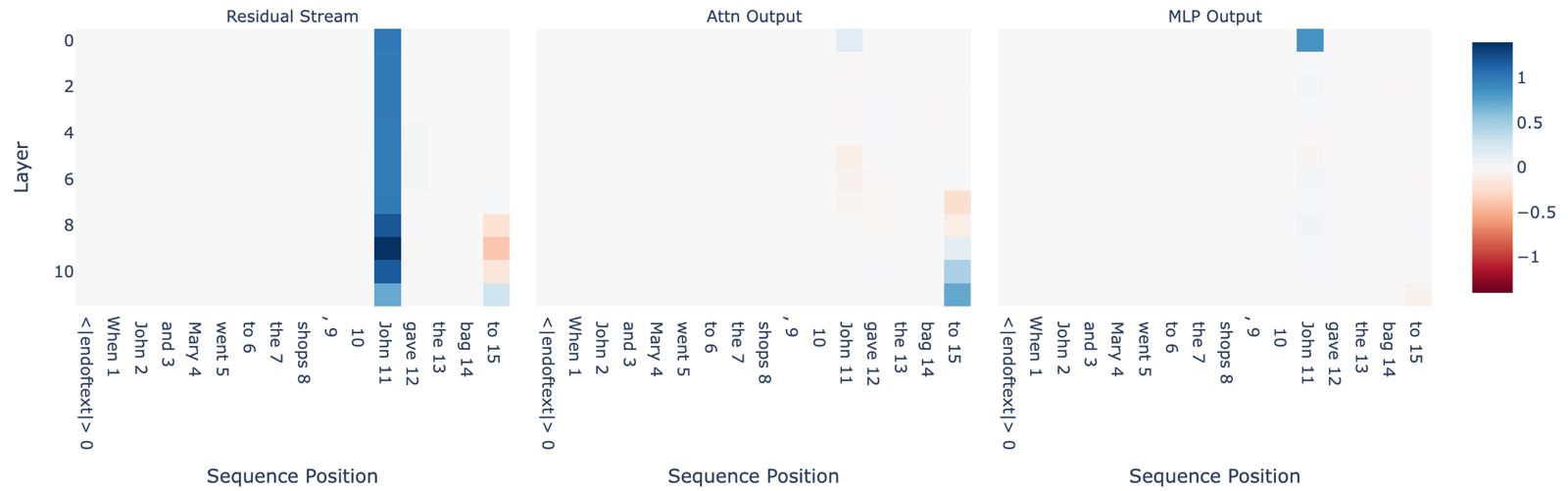}
    \caption{Component patching - obfuscated model}
    \label{fig:image_10}
\end{figure*}

When applying this component-level patching technique to the obfuscated model, a similar yet subtly different picture emerges (Figure~\ref{fig:image_10}). The general architecture of the computation—starting at the second subject token and ending at the END token—remains identifiable, but it becomes less efficient and more distributed. In particular, attention layers in layers 8, 9, and 10 assume greater importance. Notably, the patching results indicate that these layers exhibit negative contributions to the residual stream at the END token, suggesting that these components introduce transformations that degrade the clarity or fidelity of the stored information. However, this degradation is counterbalanced by positive contributions at the second subject token, implying that the model is actively encoding compensatory representations earlier in the sequence to offset later interference. The net effect is that the model’s computation is more prolonged and requires greater post hoc correction than in the standard model, confirming that obfuscation adds a layer of representational difficulty that must be managed internally.

Perhaps most striking is the reversal in attention head behavior observed in the obfuscated model. Whereas in the base model the earlier attention heads performed the bulk of meaningful computation, the obfuscated model exhibits positive contributions in the later attention heads and negative contributions in the earlier ones. This reversal suggests a redistribution of functional responsibility: because the second subject token is harder to interpret under obfuscation, earlier heads are less effective at encoding its semantics. As a result, the burden shifts to later attention heads, which must now perform the job of moving and refining the necessary information closer to the point of output generation. The indirect object still triggers attention at the END token, but this process is now deferred and spread out over a longer sequence of layers.

Finally, the behavior of the MLPs in the obfuscated model mirrors that of the base model. Once again, MLP0 emerges as the only layer of its type to exhibit significant influence, presumably continuing to function as a vital extension of the embedding. Its persistent importance across both models reinforces the hypothesis that this layer serves as an anchor point for input representation, one that later layers repeatedly refer back to for foundational semantic information.

In summary, patching at the component level confirms that attention mechanisms, especially in mid-to-late layers, carry the primary computational load for solving the indirect object identification task. While MLPs generally play a secondary role, the unique significance of MLP0 underscores the layered dependency of transformer computations. In the presence of input obfuscation, the task circuit remains intact, but it is forced to reorganize: computations that were once handled early now unfold across a broader set of layers, and the model compensates for reduced clarity by spreading effort more diffusely across attention heads in deeper blocks.

\vspace{\baselineskip}
\subsubsection{Head Patching}

To gain a deeper understanding of the internal mechanisms through which specific attention heads contribute to model behavior, we performed fine-grained head patching. This technique involves individually replacing the outputs of each attention head at specific layers and token positions—either with activations from a corrupted run (noising) or from a clean run (denoising)—to isolate their precise influence on model predictions.

\begin{figure}[h]
    \centering
    \includegraphics[width=0.4\textwidth]{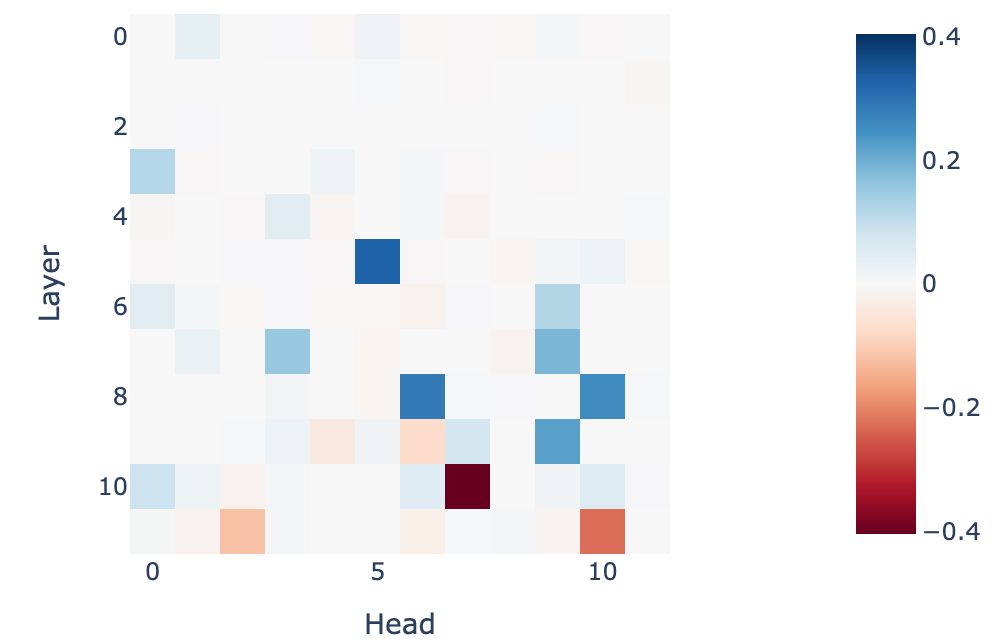}
    \caption{Head patching - base model}
    \label{fig:image_11}
\end{figure}

In the base GPT-2 model, this analysis reveals a clear functional specialization of certain heads (Figure~\ref{fig:image_11}). Specifically, attention heads 9.9 and 9.6 demonstrate strong positive contributions to the final logit difference. This means that their outputs significantly amplify the model’s confidence in selecting the correct indirect object. Their behavior suggests that they are likely performing higher-level reasoning operations, such as integrating information across sentence components, resolving coreference, or suppressing repeated token mentions—functions essential for correctly attributing the indirect object in the presence of syntactic ambiguity or duplication.

In contrast, heads 10.7 and 11.10 consistently produce negative contributions, implying that their outputs tend to reduce model accuracy for this specific task. This might indicate either interference from competing hypotheses or an overfitting to patterns that are not beneficial in the task context. Their presence suggests that late-layer heads do not uniformly contribute constructively—some may encode competing heuristics or noise that degrade performance when not properly aligned with the task objective.

Interestingly, early-layer heads, such as 3.0 and 5.5, while not contributing strongly to the final prediction, exhibit behavior consistent with primitive logical operations. These likely include low-level tasks such as detecting token boundaries, positional proximity, or duplicated token patterns. Though not directly tied to the final decision, their role is crucial in setting up the conditions for downstream attention heads to operate effectively.

\begin{figure}[h]
    \centering
    \includegraphics[width=0.4\textwidth]{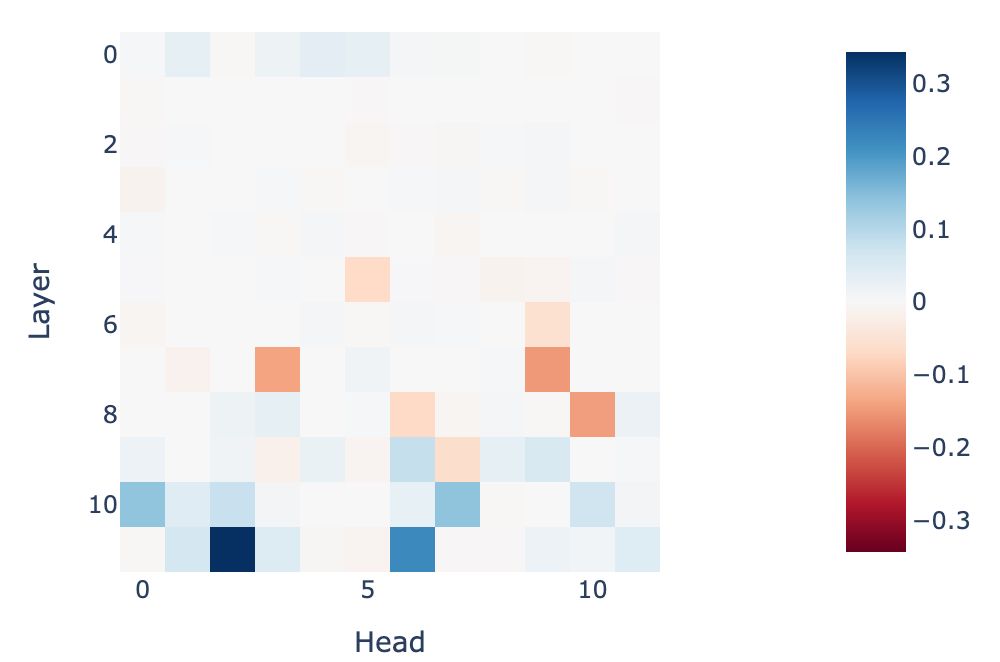}
    \caption{Head patching - obfuscated model}
    \label{fig:image_12}
\end{figure}

When the same patching analysis is conducted on the obfuscated version of the model (Figure~\ref{fig:image_12}), the observed patterns differ in notable ways. First, while the overall architecture remains intact and certain heads still exhibit discernible influence, the distribution and polarity of contributions shift. The once clearly helpful heads in layer 9, particularly 9.9 and 9.6, no longer display the same consistent positive signal. Their influence becomes more diffuse, and the clarity of their role diminishes.

Moreover, the frequency and magnitude of negative contributions increase across a broader range of heads, particularly in later layers. This suggests that obfuscation introduces noise or ambiguity into the internal representations, requiring attention heads to compensate or reconstruct more information than in the unmodified case. As a result, these heads may become overloaded or misaligned with the original circuits that were effective in the base model.

The shift in behavior supports the hypothesis that obfuscation—while preserving high-level model accuracy—alters the internal representational topology of the network. Attention heads, which in the base model act primarily as information movers (i.e., copying relevant data to critical positions), may in the obfuscated model take on more processing-intensive roles. That is, instead of just relocating salient features, they appear to transform and reinterpret them in a way that is compatible with downstream expectations, potentially compensating for the distortion introduced by the encrypted or transformed inputs.

Overall, this analysis highlights that obfuscation does not eliminate task-relevant circuits, but reshapes their internal dynamics. The altered behavior of attention heads suggests a reorganization of functional responsibilities across the model, reaffirming the importance of interpretability methods such as patching to uncover how robust and adaptive transformer-based models are when confronted with representational distortions.

\vspace{\baselineskip}
\section{Conclusion}
\vspace{\baselineskip}

This study set out to determine whether architectural obfuscation—advertised as a low-overhead privacy defence—destroys the ability of modern mechanistic tools to explain a Transformer or merely shifts that explanatory task into a different coordinate system.  By applying residual-stream patching, block-component patching and head patching to GPT-2-small before and after a representative obfuscation, we obtained a clear picture.  The essential circuit that moves information from the duplicated-subject token to the END token survives every transformation; however, activation patterns become markedly more diffuse, attention roles invert in the later blocks, and head-level importance is scattered across a broader set of neurons.  These changes make causal tracing noisier and prompt reconstruction far less precise while leaving top-line accuracy intact.

From a mechanistic-interpretability perspective, the results validate obfuscation as a useful—but not absolute—privacy layer.  It hinders fine-grained reverse engineering without erasing the coarse computations that safety or compliance audits may still need to inspect.  Having established that lightweight scrambling offers real benefits, the next step is to strengthen security guarantees without forfeiting interpretability.  Cryptographic techniques are the natural frontier here.  Fully homomorphic encryption is especially promising: it processes ciphertext end-to-end, neutralising the leakage channels that permutation schemes cannot close, yet recent progress in diagonal packing and low-depth polynomial approximations suggests that the latency penalty is no longer prohibitive for medium-size models.  Bridging architectural obfuscation with homomorphic or hybrid MPC–HE protocols could yield systems that are simultaneously private, auditable and efficient.  In short, the road ahead lies in marrying the interpretability-preserving qualities of obfuscation with the provable security of modern cryptography.

\printbibliography

\end{document}